\documentclass[12pt,twoside]{article}
\usepackage{fleqn,espcrc1}

\usepackage{graphicx}
\usepackage[figuresright]{rotating}

\title{The quark-photon vertex and
meson electromagnetic form factors}

\author{P. Maris and P.C. Tandy\address{Dept. of Physics, 
        Kent State University, Kent OH 44242}%
        \thanks{This work was funded by the National Science Foundation 
        under grant No.~PHY97-22429, and benefited from the resources of 
        the National Energy Research Scientific Computing Center.}}

\begin{document}

\maketitle

\begin{abstract}
The ladder Bethe--Salpeter solution for the dressed photon-quark vertex
is used to study the low-momentum behavior of the pion electromagnetic
and the \mbox{$\gamma^\star \pi^0 \gamma$} transition form factors.
With model parameters previously fixed by light meson masses and decay
constants, the low-momentum slope of both form factors is in excellent
agreement with the data.  In comparison, the often-used Ball--Chiu
Ansatz for the vertex is found to be deficient; less than half of the
obtained $r_\pi^2$ is generated by that Ansatz while the remainder of
the charge radius could be attributed to the tail of the $\rho$
resonance.
\end{abstract}

\section{DYSON--SCHWINGER EQUATIONS}

The Dyson--Schwinger equations [DSEs] form a useful tool for
nonperturbative QCD modeling of hadrons and their interactions.  They
have been successfully applied to calculate properties of light vector
and pseudoscalar mesons~\cite{MR97,MT99}, as described elsewhere in
these proceedings~\cite{PM99}.  The dressed-quark propagator, as
obtained from its DSE, together with the Bethe--Salpeter amplitude as
obtained from the Bethe--Salpeter equation [BSE] for $q\bar q$ bound
states, form essential ingredients for calculations of meson couplings
and form factors~\cite{PCT97}.  To describe electromagnetic interactions
of hadrons, we also need the nonperturbatively dressed quark-photon
vertex.  Here, we use a solution of the DSE for the quark-photon vertex
under the same truncation as used in Refs.~\cite{MR97,MT99,PM99} for
light mesons.

The DSE for the renormalized dressed-quark propagator in Euclidean space
is
\begin{eqnarray}
\label{gendse}
 S(p)^{-1} & = & Z_2\,i\gamma\cdot p + Z_4\,m(\mu)
        + Z_1 \int \frac{d^4q}{(2\pi)^4} \,g^2 D_{\mu\nu}(p-q) 
        \frac{\lambda^a}{2}\gamma_\mu S(q)\Gamma^a_\nu(q,p) \,,
\end{eqnarray}
where $D_{\mu\nu}(k)$ is the dressed-gluon propagator and
$\Gamma^a_\nu(q;p)$ the dressed-quark-gluon vertex.  The solution of
Eq.~(\ref{gendse}) has the form \mbox{$S(p)^{-1} = i \gamma\cdot p
A(p^2) + B(p^2)$} and is renormalized at spacelike $\mu^2$ according to
\mbox{$A(\mu^2)=1$} and \mbox{$B(\mu^2)=m(\mu)$} with $m(\mu)$ the
current quark mass.

The DSE for the quark-photon vertex \mbox{$\tilde{\Gamma}_\mu(p;Q)=
\hat{Q} \Gamma_\mu(p;Q)$} is the inhomogeneous BSE
\begin{equation}
\tilde{\Gamma}_\mu(p;Q) = Z_2\; \hat{Q}\; \gamma_\mu + 
        \int^\Lambda\!\!\frac{d^4q}{(2\pi)^4} \; K(p,q;Q) 
        S(q+Q/2)\tilde{\Gamma}_\mu(q;Q) S(q-Q/2) \,,
\label{verBSE}
\end{equation}
where $\hat{Q}$ is the charge operator.  The kernel $K$ is the
renormalized, amputated $\bar q q$ scattering kernel that is irreducible
with respect to a pair of $\bar q q$ lines.  Solutions of the
homogeneous version of Eq.~(\ref{verBSE}) define the vector meson bound
states at \mbox{$Q^2=-m^2$}.  It follows that $\tilde{\Gamma}_\mu(p;Q)$
has poles at those locations.

We use a ladder truncation for the BSE in conjunction with a rainbow
truncation \mbox{$\Gamma^a_\nu(q,p) \rightarrow \gamma_\nu \lambda^a/2$}
for the quark DSE.  Both the vector Ward--Takahashi identity [WTI] for
the quark-photon vertex and the axial-vector WTI are preserved in this
truncation.  This ensures both current conservation and the existence of
massless pseudoscalar mesons if chiral symmetry is broken dynamically:
pions are Goldstone bosons~\cite{MR97}.

The details of the model can be found in Refs.~\cite{MT99,PM99}.  It
leads to chiral symmetry breaking and confinement; furthermore, at large
momenta, our effective interaction reduces to the perturbative running
coupling and thus preserves the one-loop renormalization group behavior
of QCD and reproduces perturbative results in the ultraviolet region.
The model gives a good description of the $\pi$, $\rho$, $K$, $K^\star$
and $\phi$ masses and decay constants~\cite{MT99,PM99}.

\section{THE DRESSED QUARK-PHOTON VERTEX}

The general form of the quark-photon vertex $\Gamma_\mu(q;Q)$ can be
decomposed into twelve independent Lorentz covariants.  Four of these
covariants, representing the longitudinal components, are uniquely
determined by the vector WTI
\begin{equation}
i\,Q_\mu \,\Gamma_{\mu}(p;Q)  =  S^{-1}(p+Q/2) - S^{-1}(p-Q/2) \;.
\label{wtid}
\end{equation}
The eight transverse components of $\Gamma_\mu(p;Q)$ are not constrained by
the WTI, except at $Q=0$, where the WTI reduces to \mbox{$i
\,\Gamma_{\mu}(p;0)=$} \mbox{$\partial S(p)^{-1}/\partial p_\mu$}.

It is obvious from Eq.~(\ref{wtid}) that the bare vertex $\gamma_\mu$ is
a bad approximation if the quark self-energy is momentum dependent as is
realistically the case.  For QCD modeling of electromagnetic coupling to
hadrons, it has been common practice~\cite{MR98,PCT99} to avoid a
numerical study of the quark-photon vertex, and use the so-called
Ball--Chiu [BC] Ansatz~\cite{BallChiu}, which expresses the vertex in
terms of the dressed-quark propagator functions $A$ and $B$
\begin{equation}
 \Gamma_{\mu}^{BC}(p;Q)  =
       \gamma_\mu\,\frac{A(p^2_+) +A(p^2_-)}{2}
        + 2\,(\gamma\cdot p)\,p_\mu\,
        \frac{A(p^2_+) -A(p^2_-)}{p_+^2 - p^2_-}
        - 2\,i\,p_\mu \, \frac{B(p^2_+) -B(p^2_-)}{p_+^2 - p^2_-} \, ,
\label{bcver}
\end{equation}
where \mbox{$p_\pm = p \pm Q/2$}. This satisfies the WTI,
Eq.(\ref{wtid}), transforms correctly under CPT, and has the correct
perturbative limit $\gamma_\mu$ in the extreme ultraviolet.  The
longitudinal components of $\Gamma_\mu^{BC}$ are exact, but the
transverse components are correct only at \mbox{$Q=0$}.  In particular,
$\Gamma_\mu^{BC}$ does not have the vector meson poles.  This should be
of little concern for form factors at large spacelike $Q^2$; however,
for \mbox{$Q^2 \approx 0$} the situation is less clear.

Our numerical solution of Eq.~(\ref{verBSE}) shows clearly the vector
meson pole in all eight transverse amplitudes.  The solution for the
four longitudinal amplitudes agrees perfectly with the BC Ansatz, as
required by the WTI.  Our transverse solution agrees with the BC Ansatz
only at spacelike asymptotic momenta.  At low $Q^2$ it departs
significantly from this Ansatz; although there is necessarily agreement
at the point
\mbox{$Q=0$}, the $Q$-dependence of the DSE solution is much larger than
that of the BC Ansatz.  Near the $\rho$ pole, the quark-photon vertex
behaves like
\begin{equation}
\tilde{\Gamma}_\mu(p;Q) \;\simeq\;      \frac{\Gamma_\mu^\rho(p;Q) 
                          m_\rho^2/g_\rho }{Q^2 + m_\rho^2}     \,,
\label{verres}
\end{equation}
where the $\rho-\gamma$ coupling strength \mbox{$m^2_\rho/g_\rho$}
associated with the \mbox{$\rho \rightarrow e^+ \, e^-$} decay is well
reproduced by the present model~\cite{MT99,PM99}.  There is no unique
decomposition of the vertex into resonant and non-resonant terms away
from the pole, but over a limited interval near \mbox{$Q^2 \approx 0$}
the difference between the DSE solution and the BC Ansatz can be
approximated by Eq.~(\ref{verres}) with \mbox{$m^2_\rho \rightarrow
-Q^2$} in the numerator, which one can call the tail of the $\rho$
resonance.  However, the DSE solution for the vertex is the appropriate
representation containing both the resonant and non-resonant parts of
the vertex.

\section{MESON FORM FACTORS}

In the impulse approximation, both the pion charge form factor and the
$\gamma^\star\,\pi\,\gamma$ transition form factor are described by a
triangle diagram, with one or two pion Bethe--Salpeter amplitudes, one
or two quark-photon vertices, and three dressed-quark propagators.  We
obtain these elements from the appropriate BSE and DSE, within the same
model.  We compare the form factor results using: 1) a bare vertex; 2)
the BC Ansatz; 3) the DSE solution for the quark-photon vertex.  Note
that the WTI ensures electromagnetic current conservation in both 2) and
3), but not in approximation 1), which violates the WTI.

The impulse approximation for the pion form factor gives
\begin{equation}
\label{gpp}
 F_\pi(Q^2) P_\nu = 
        N_c \int\!\frac{d^4q}{(2\pi)^4} {\rm Tr}\left[
        \Gamma_\pi(k_+;-P_+)\, S(q_{+-})\,i\Gamma_\nu(q_+;Q)\, 
        S(q_{++}) \, \Gamma_\pi(k_-;P_-)\, S(q_-)\right]\,,
\end{equation}
where $Q$ is the photon momentum, \mbox{$P_\pm = P \pm Q/2$},
\mbox{$q_\pm = q \pm P/2$}, \mbox{$q_{+ \pm} = q_+ \pm Q/2$}
and \mbox{$k_\pm = q \pm Q/4$}.  It is evident that the $Q^2$ dependence
of $F_\pi$ comes from both the quark substructure of the pion and the
$Q$-dependence of the quark-photon vertex.  Due to Eq.~(\ref{verres}),
$F_\pi(Q^2)$ will exhibit a resonance peak at timelike momenta $Q^2$
near $-m_\rho^2$.

A long-standing issue in hadronic physics is the question of the extent to
which $F_\pi(Q^2)$ at low spacelike $Q^2$ can still be described by the
$\rho$ resonance mechanism.  This is an essential element of vector
meson dominance (VMD), which leads to~\cite{VMD}
\begin{equation}
{}F^{VMD}_\pi(Q^2) = 1 - \frac{ g_{\rho\pi\pi} \; Q^2} 
                        {g_\rho \, (Q^2 + m_\rho^2) } \;.
\label{piffvmd}
\end{equation}
The first term arises from the non-resonant photon coupling to a point
pion; the only $Q^2$ dependence in VMD comes from the resonant mechanism
with the produced $\rho$ having a point coupling to the pion.  The pion
charge radius \mbox{$r_\pi^2 = $} \mbox{$-6 F^\prime_\pi(0)$} thus
becomes \mbox{$ 6 g_{\rho \pi\pi}/(m_\rho^2 g_\rho) \sim 0.48~{\rm
fm}^2$} which compares favorably with the experimental value $0.44~{\rm
fm}^2$.

\begin{figure}[htb]
\vspace*{4mm}
\begin{minipage}[t]{160mm}
\includegraphics[height=14pc]{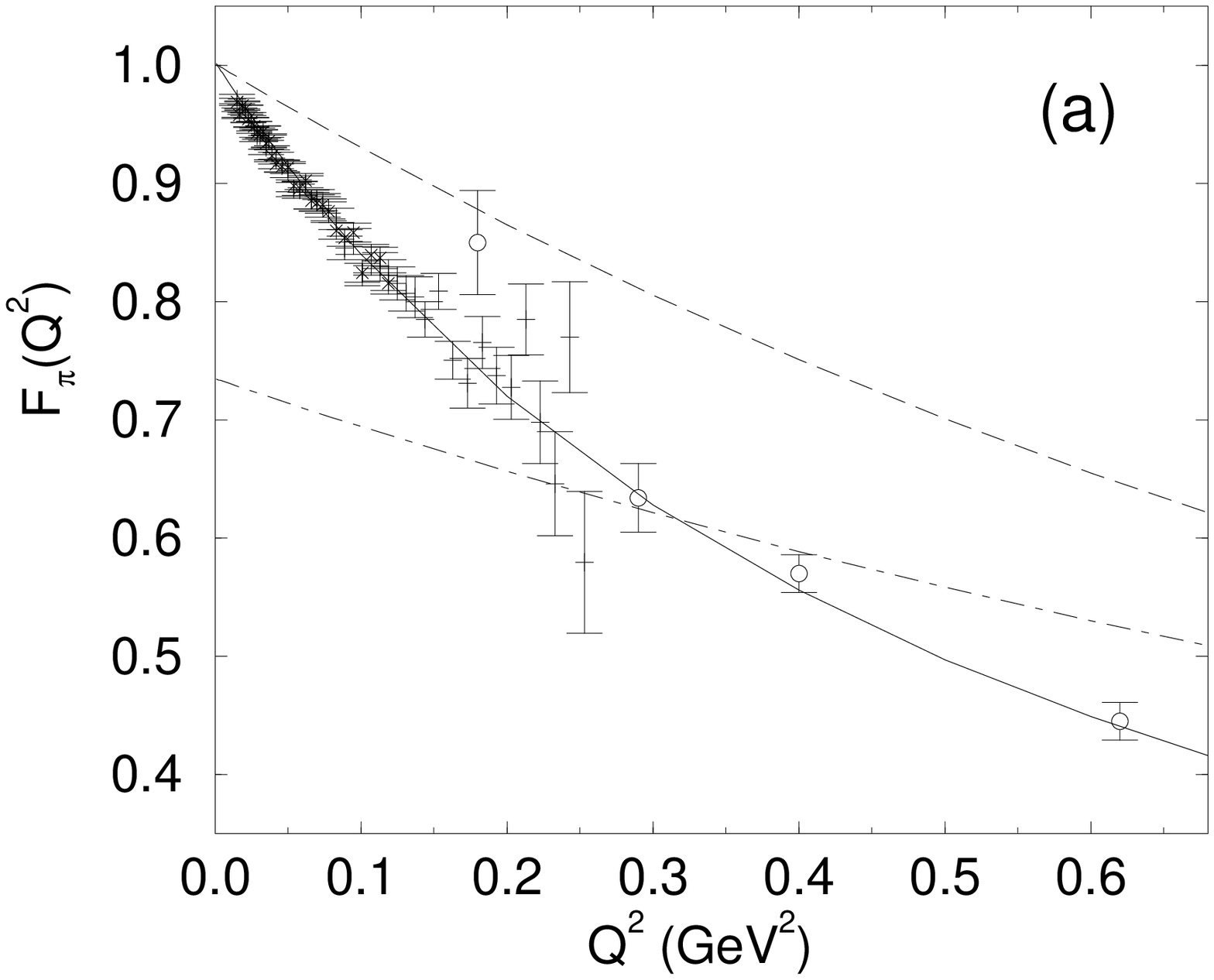}
\hspace{\fill}
\includegraphics[height=14pc]{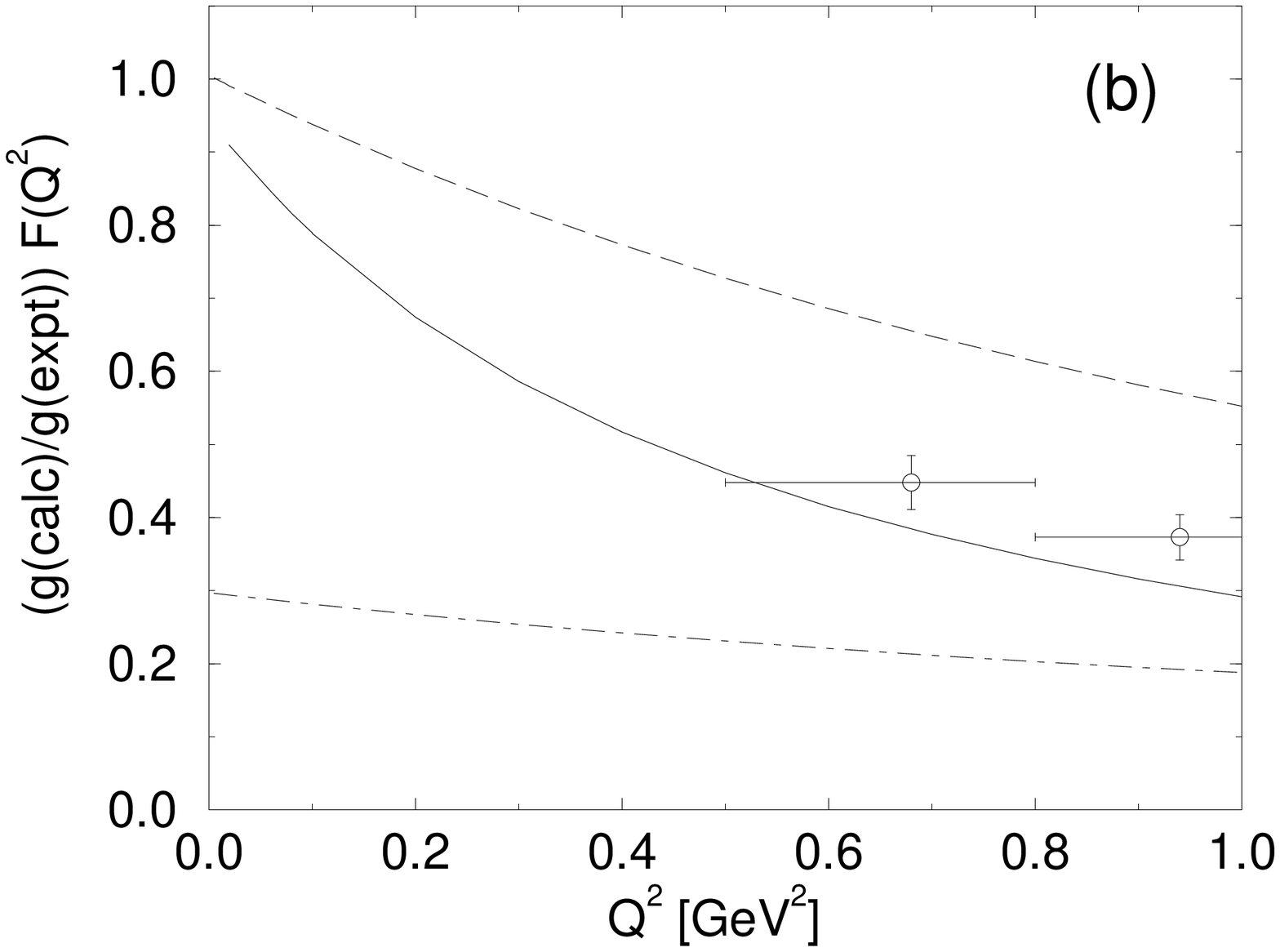}
\vspace*{-8mm}
\caption{
The pion form factor (a) and \protect{$\gamma^\star\pi\gamma$}
transition form factor (b) using a bare vertex (dot-dash), the BC Ansatz
(dashed) and our numerical solution of Eq.~(\ref{verBSE}) (solid).}
\label{Fig:ffs}
\end{minipage}
\end{figure}

In Fig.~\ref{Fig:ffs}(a) we show our results from Eq.~(\ref{gpp}).
Clearly a bare vertex is incorrect: current conservation, which ensures
\mbox{$F_\pi(0)=1$}, is violated.  Use of the BC Ansatz conserves the
current; but the resulting $r^2_\pi = 0.18~{\rm fm}^2$ is less than half
the experimental value and the curve misses the data completely.  The
DSE solution for the vertex agrees very well with the data and produces
$r^2_\pi = 0.45~{\rm fm}^2$, {\em without fine tuning the model
parameters}: the parameters are completely fixed in
Refs.~\cite{MT99,PM99}.  This indicates that as much as half of
$r_\pi^2$ can be attributed to a reasonable extrapolation of the $\rho$
resonance mechanism.  On the other hand, the strict VMD picture is too
simple; at least 40\% of $r_\pi^2$ arises from the non-resonant photon
coupling to the quark substructure of the pion.

The impulse approximation for the $\gamma^\star\pi\gamma$ vertex 
with $\gamma^\ast$ momentum $Q$ is
\begin{eqnarray}
\label{int}
\lefteqn{ \Lambda_{\mu\nu}(P,Q)=i\frac{\alpha }{\pi f_{\pi }}
        \,\epsilon_{\mu \nu \alpha \beta }\,P_{\alpha }Q_{\beta }
        \, g_{\pi\gamma\gamma}\,F(Q^2)  } \\
& & \nonumber
        =\frac{N_c}{3}\, \int\!\frac{d^4k}{(2\pi)^4}
        {\rm Tr}\left[S(q^\prime)\, i\Gamma_\nu (k^\prime;Q) 
        \,S(q^{\prime\prime})\, i\Gamma_\mu (k^{\prime\prime};-P-Q)
        \, S(q^{\prime\prime\prime})
        \,\Gamma_\pi (k;P)\right] \;.
\end{eqnarray}
where the momenta follow from momentum conservation.  In the chiral
limit, the value at $Q^2 = 0$, corresponding to the decay \mbox{$\pi^0
\rightarrow \gamma\gamma$}, is given by the axial anomaly and its value
\mbox{$g^{0}_{\pi\gamma\gamma}=1/2$} is a direct consequence of only
gauge invariance and chiral symmetry; this value corresponds well with
the experimental width of $7.7~{\rm eV}$.  In Fig.~\ref{Fig:ffs}(b) we
show our results, normalized to the experimental $g_{\pi\gamma\gamma}$.
A bare vertex does not reproduce the anomaly since it violates WTIs.
Both the BC Ansatz and the DSE vertex solution reproduce the anomaly
value, but the BC Ansatz overestimates the form factor at small but
nonzero spacelike momenta and gives an interaction radius $r^2 =
0.13~{\rm fm}^2$, compared to the experimental value $r^2
\sim 0.42~{\rm fm}^2$~\cite{gpgexp}. The vertex DSE solution gives
results remarkably close to the data and $r^2 = 0.40~{\rm fm}^2$, again
indicating that the BC Ansatz underestimates the $Q^2$ dependence of the
form factor, related to the absence of a $\rho$ resonance.

\end{document}